\documentclass[MSNbibl,nameyear,dvips]{arxstspdf}
\usepackage{flushend}
\usepackage{stfloats}

\volume{26}
\issue{2}
\pubyear{2011}
\firstpage{185}
\lastpage{186}
\referstodoi{10.1214/10-STS318}
\doi{10.1214/10-STS318REJ}

\begin{document}
\begin{frontmatter}

\title{Rejoinder}
\runtitle{Rejoinder}

\begin{aug}
\author[a]{\fnms{Roderick} \snm{Little}\corref{}\ead[label=e1]{rlittle@umich.edu}}

\runauthor{R. Little}

\affiliation{University of Michigan}

\address[a]{Roderick Little is Richard D. Remington Collegiate Professor, Department of
Biostatistics, University of Michigan, 1415 Washington Heights, Ann Arbor, Michigan 48109, USA. \printead{e1}.}

\end{aug}

% ABSTRACT

% KEYWORDS

\end{frontmatter}

I thank Michael Larsen and Nathaniel Schenker for their thoughtful
contributions, which usefully reinforce and expand my arguments.

Any differences between my perspective and that described by Schenker
are minor. He defends a pragmatic approach to the Bayesian/frequentist
divide, but in my applied work I confess I am pragmatic too. I provide
confidence intervals and even $P$ values to my biomedical collaborators,
rather than posterior credibility intervals. I fear these would meet
with head-scratching (not to mention article rejection) given current
conventions of statistical reporting in medical journals. Like Schenker,
I rely on similarities of frequentist and Bayesian interval estimates in
many standard models. I~am more focused on developing good scientific
models than on elicitation of prior distributions.

Despite concessions to current-day realities, thinking about and
articulating the underlying principles that should guide our methodology
is worthwhile. For example, in the survey sampling setting, I do not
believe design-based inference is appropriate for some problems, and
model-based inference is appropriate for others.

Larsen calls for more examples of how to achieve Bayesian calibration in
practice. A very reasonable request, but complete textbooks are needed
to do any kind of justice to that aim; my examples barely scratch the
surface. Concerning his specific comments, I think Bayesian hierarchical
models are huge\-ly valuable; model checks seeking a good fit to the
observed data are important, but unfortunately not sufficient to
guarantee good predictions for missing data and target unknowns. A
calibrated Bayes perspective would I hope push the field toward more
research and training on how to develop good models in practice.

Schenker's theoretical and applied work on multiple imputation is
influential, and the examples cited in his discussion add real substance
to my musings. Larsen's multiple imputation example allows me to
illustrate Schenker's comment in his Section 4.4 on the danger of
omitting important variables in the imputation model. The relationship
between parental longevity and offspring's diabetes was under study, and
age of death was missing for parents currently alive. Larsen's
imputation model conditioned on diabetes status of the offspring. This
is important, since a multiple imputation model for this variable that
failed to condition on offspring's diabetes status would lead to
attenuation of the estimated relationship between these variables---indeed, it might well be worse than discarding the incomplete cases,
which would distort the distribution of parental age at death, but not
necessarily the relationship under study.

Both discussants consider the Bayes/frequentist divide in the context of
inference from survey samples. As Schenker states, the debate is
particularly lively in that area, given that the prevailing philosophy
is to base inference on the randomization distribution that governs
sample selection. My perspective is described in Little (\citeyear{l2004}), but let
me respond to some of Larsen's comments. The goal of
\textit{design-based} survey inference is ``frequentist in nature,'' but
for me (well-calibrated), Bayesian inference is just as useful and
appropriate for inference about finite population quantities as it is
for model parameters. The bald statement that ``no model at all is
involved in design-based inference'' is oversimplified, since (as Larsen
points out) design-based inference without any consideration of the
implicit model underlying the choice of estimator leads to absurdities
like Basu's (\citeyear{Bas71}) famous elephant example.

Terminology can be confusing, and Larsen's comment allows me to draw
distinctions between my use of the term ``calibration'' and other uses.
Deville and S\"{a}rndal (\citeyear{ds1992}) discuss calibration of estimates to
aggregate statistics. This form of data calibration is (in principle)
automatically achieved by a Bayesian model for prediction that
incorporates this information, without the need for Deville and
S\"{a}rndal's ad-hoc distance measure, though achieving it exactly may
be challenging. Larsen also mentions ``broader modeling options'' in
S\"{a}rndal, Swensson and Wretman (\citeyear{ssw1992}). These model-assisted methods
``calibrate'' model predictions with design-weighted adjustments based
on the model residuals. They lead to compromises between model
predictions and direct design-based estimates, similar to the doubly
robust competitors to the PPSP\vadjust{\goodbreak} method in my Example 4. In my view, these
approaches, which are ultimately design-based, are inferior to judicious
application of the calibrated Bayes approach. For simulation evidence in
the context of my Example 4, see Kang and Schafer (\citeyear{KanSch07}) and Zhang and
Little (\citeyear{ZhaLit}).

In my discussion of Hansen, Madow and Tepping (\citeyear{hmt1983}), I opined that the
way to mitigate the effects of model misspecification is not to modify
the estimator, as in these model assisted methods, but to modify the
model. I have not changed my opinion.%\

% imsref loaded by arune.pranskunaite, 2011-01-28 16:20:10

\end{document}